\documentclass[aps,superscriptaddress,showpacs]{revtex4}
\usepackage{amssymb}
\usepackage{amsmath}
\usepackage{graphicx}
\usepackage[normalem]{ulem}

\begin{document}

\title{Breaking the EOS-Gravity Degeneracy with Masses and Pulsating Frequencies of Neutron Stars}
\author{Weikang Lin\footnote{Current address: Department of Physics, The University of Texas at Dallas, Richardson, TX 75080, USA}}
\affiliation{Department of Physics and Astronomy, Texas A\&M University-Commerce, Commerce, Texas 75429-3011, USA}
\author{Bao-An Li\footnote{Corresponding author, Bao-An.Li$@$Tamuc.edu}}
\affiliation{Department of Physics and Astronomy, Texas A\&M
University-Commerce, Commerce, Texas 75429-3011, USA}
\author{Lie-Wen Chen}
\affiliation{Department of Physics and Astronomy and Shanghai Key Laboratory for Particle Physics and Cosmology, Shanghai Jiao Tong University, Shanghai 200240, China}
\affiliation{Center of Theoretical Nuclear Physics, National Laboratory of Heavy Ion Accelerator, Lanzhou 730000, China}
\author{De-Hua Wen}
\affiliation{Department of Physics, South China University of
Technology,Guangzhou 510641, P.R. China}
\author{Jun Xu}
\affiliation{Shanghai Institute of Applied Physics, Chinese Academy of Sciences, Shanghai 201800, China}
\date{\today}

\begin{abstract}
A thorough understanding of many astrophysical phenomena associated with compact objects
requires reliable knowledge about both the equation of state (EOS)
of super-dense nuclear matter and the theory of strong-field gravity
simultaneously because of the EOS-gravity degeneracy.
Currently, variations of the neutron star (NS) mass-radius correlation
from using alternative gravity theories are much larger than those
from changing the NS matter EOS within known constraints.
At least two independent observables are required to break the
EOS-gravity degeneracy. Using model EOSs for hybrid stars and a
Yukawa-type non-Newtonian gravity, we investigate both the mass-radius correlation and pulsating frequencies of NSs.
While the maximum mass of NSs increases, the frequencies of the
$f$, $p_1$, $p_2$, and $w_I$ pulsating modes are found to decrease
with the increasing strength of the Yukawa-type non-Newtonian gravity, providing a useful reference for future determination
simultaneously of both the strong-field gravity and the supranuclear
EOS by combining data of x-ray and gravitational wave emissions of neutron stars.
\end{abstract}

\pacs{26.60.-c, 97.60.Jd, 14.70.Pw}

\keywords{Equation of State, dense matter, neutron stars, gravity}
\maketitle \maketitle

\section{Introduction}
Neutron stars (NSs) are natural testing grounds of grand unification
theories of fundamental forces as their compositions are determined
mainly by the weak and electromagnetic forces through the $\beta$
equilibrium and charge neutrality conditions while their mechanical
stability is maintained by the balance between the strong and
gravitational forces. Moreover, neutron stars are among the densest
objects with the strongest gravity in the Universe, making them
ideal places to test simultaneously our understanding about the
nature of super-dense nuclear matter and the strong-field
predictions of general relativity (GR). Many fundamental questions about both
the strong-field gravity and properties of super-dense nuclear matter remain to be
answered \cite{quet}.
Tied to the fundamental strong equivalence principle \cite{Yun10}, the degeneracy
between the NS matter content and its gravity requires a simultaneous
understanding about both the strong-field gravity and the equation of state (EOS) of super-dense nuclear matter.
For example, the masses and radii of neutron stars are determined by the
balance between the strong-field behavior of gravity and the
EOS of dense nuclear matter. While the nature of
super-dense matter in NSs and its EOS are currently largely unknown,
there is also no fundamental reason to choose Einstein's GR over
other alternatives and it is known that the GR theory itself may
break down at the strong-field limit \cite{Psa08}.
Indeed, it is very interesting to note that alternative gravity theories, which have
all passed low-field tests but diverge from GR in the strong-field
regime, predict neutron stars with significantly different properties
than their GR counterparts \cite{Yun10}. In fact, the variations in
the predicted NS mass-radius correlation from modifying gravity are
larger than those from varying the EOS of super-dense matter in neutron stars \cite{Ded03}.
Therefore, the EOS of super-dense matter and the strong-field gravity have to be studied
simultaneously to understand properties of neutron stars accurately. Moreover,
the gravity-EOS degeneracy can only be broken by using at least two independent observables.
We notice that effects of modified gravity on properties of
neutron stars have been under intense investigation. The conclusions are, however,  strongly
model-dependent, see, e.g., Refs. \cite{Ger01,Wis02,Aza08,Kri09,Wen0911,Ova10}.
In particular, it is interesting to mention that the very recent test of strong-field gravity
using the binding energy of the most massive NS observed so far, the PSR J0348+0432 of mass $2.01\pm0.04M_{\odot}$,
found no evidence of any deviation from the GR prediction \cite{PSR348}. However,
it is worth noting that this analysis used a very stiff EOS with an incompressibility of $K_0=546$ MeV
at normal nuclear matter density \cite{Ha81,Serot79}. Such a super-stiff incompressibility is more than
twice the value of $K_0=240 \pm 20$ MeV extracted from analyzing various experiments
in terrestrial nuclear laboratories during the last 30 years \cite{Garg}.

Many kinds of modified gravity theories have been published ever since Einstein published his GR in 1915,
see, e.g., refs. \cite{Cap13,Ova13} and references therein for a very recent reviews.
Besides the Newtonian potential, an extra Yukawa term naturally arise in the gravitational potential at the weak-field limit
in some of these theories, e.g., f(R) \cite{fr1,fr2}, the nonsymmetric gravitational theory (NGT) \cite{Mof96} and Modified Gravity (MOG) \cite{mog},
see ref. \cite{NAP} for a very recent review. Such theories are able to explain successfully the
well-known astrophysical observations, such as the flat galaxy rotation
curves \cite{Mof96,San84} and the Bullet Cluster 1E0657-558 observations~\cite{Bro07}
in the absence of dark matter. Basically, this is because particles with masses, besides
the presumably massless graviton, may also carry the gravitational force \cite{San84}.
Following Fujii \cite{Fujii71,Fuj2,Fujii91}, we model effects of the
modified gravity by adding a Yukawa term to the conventional gravitational potential.

To our best knowledge, while various properties of neutron stars and gravitational waves
have been used to probe the EOS and gravity often individually, it is not so clear what
observables should be used to effectively break the EOS-gravity
degeneracy. In this work, we show that the simultaneous measurement of the NS mass and its pulsating
frequencies is useful for breaking the EOS-gravity degeneracy. Model
EOSs for super-dense matter with or without the hadron-quark phase
transition are considered. For hybrid stars, the EOS is constructed
by using an extended isospin- and momentum-dependent nuclear
effective interaction for the hyperonic phase, the MIT bag model for
the quark phase, and the Gibbs conditions for the mixed phase.
Model parameters are constrained by existing experimental data whenever possible in
constructing the EOS. By turning on or off the hadron-quark phase
transition, varying the MIT bag constant or the strength of the
Yukawa term, we explore the possibility of breaking the EOS-gravity
degeneracy by using the mass and the frequencies of the $f$, $p_1$,
$p_2$, and $w_I$ pulsating modes of neutron stars. Since effects of
the EOS on properties of both neutron stars and gravitational waves are
better known as a result of extensive previous studies in the
literature, we focus on effects of the Yukawa-type non-Newtonian
gravity on both the maximum mass and pulsating frequencies of
neutron stars. While the NS maximum mass is found to increase, the
frequencies of the $f$, $p_1$, $p_2$, and $w_I$ pulsating modes
decrease with the increasing strength of the Yukawa term. A
simultaneous measurement of the NS mass and the frequency of one of
its pulsating modes, preferably the $f$-model, can thus readily
break the EOS-gravity degeneracy, providing a useful reference for
testing simultaneously both the strong-field gravity and supranuclear EOS
using x-rays and gravitational waves hopefully in
the near future.

\section{Yukawa-type Non-Newtonian gravity and model EOS of hybrid stars}
\begin{figure}
\begin{center}
\includegraphics[width=8cm,height=6.5cm]{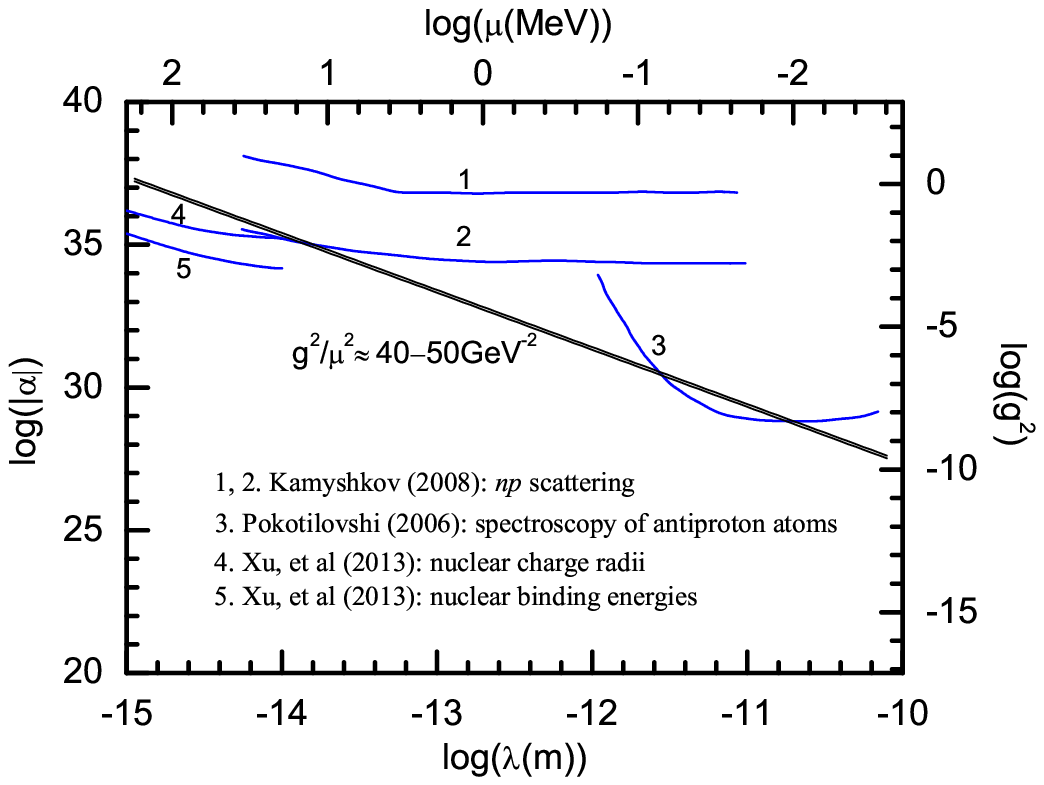}
\includegraphics[width=8cm,height=6.5cm]{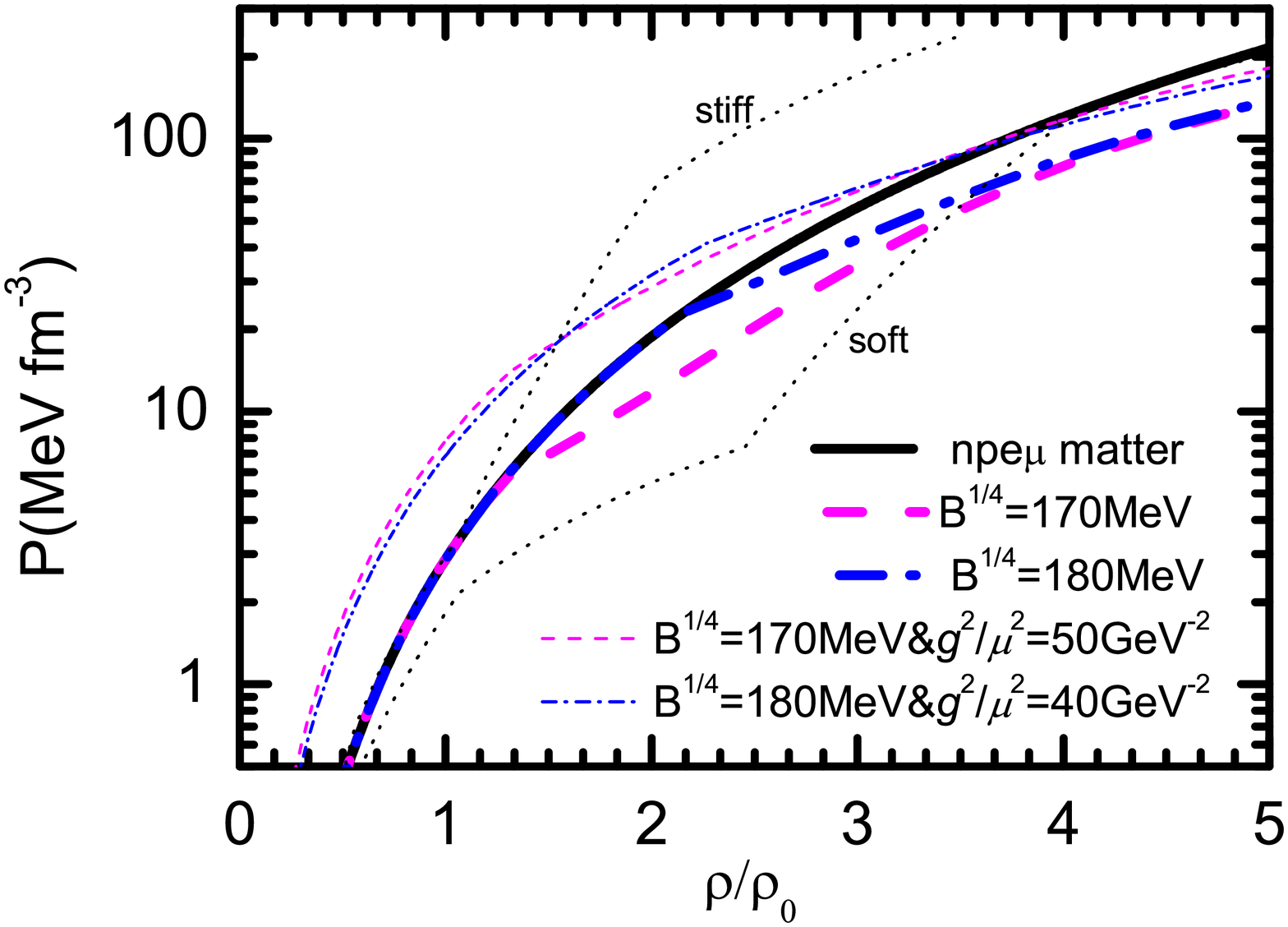}
\caption{\label{fig1}  (Color online) Left: Constraints on the
strength and range of the Yukawa term from terrestrial nuclear
experiments in comparison with $g^{2}/\mu^{2}\approx 40-50$
GeV$^{-2}$; Right: Model EOSs for hybrid stars with and without the
Yukawa contribution using MIT bag constant $B^{1/4}$=170 MeV and
$B^{1/4}$=180 MeV, respectively. The soft and stiff EOSs for $npe$ matter are taken from refs. \cite{Lat12,Heb13}.}
\end{center}
\end{figure}

Despite the fact that gravity is the first force discovered in
nature, the quest to unify it with other fundamental forces remains
elusive because of its apparent weakness at short distance, see,
e.g., Refs. \cite{Ark98,Pea01,Hoy03,Long03,Jean03,Boehm04a,Boe04,Dec05}. In
developing grand unification theories, the conventional
inverse-square-law (ISL) of Newtonian gravitational force has to be
modified due to either the geometrical effect of the extra
space-time dimensions predicted by string theories and/or the
exchange of weakly interacting bosons, such as the neutral spin-1
vector $U$-boson \cite{Fayet}, proposed in the super-symmetric
extension of the Standard Model, see, e.g., Refs.
\cite{Adel03,Fis99,New09,Uzan03,Rey05} for recent reviews. As an Ansatz, Fujii
\cite{Fujii71,Fujii91} first proposed that the non-Newtonian gravity
can be described by adding a Yukawa term to the conventional
gravitational potential between two objects of mass $m_1$ and $m_2$,
i.e.,
\begin{equation}
V(r)=-\frac{Gm_{1}m_{2}}{r}(1+\alpha e^{-r/\lambda}),
\end{equation}
where $\alpha$ is a dimensionless strength parameter, $\lambda$ is
the length scale, and $G$ is the gravitational constant. In the
boson exchange picture, the strength of the Yukawa potential between
two baryons can be expressed as $\alpha=\pm g^2/(4\pi Gm_b^{2})$
where $\pm$ stands for scalar/vector bosons, $m_b$ is the baryon
mass, and $\lambda=1/\mu$ (in natural units), with $g$ and $\mu$
being the boson-baryon coupling constant and the boson mass,
respectively. As mentioned earlier, the Yukawa term appears naturally at the weak-field limit
in several modern modified gravity theories and has been used to explain successfully
the flat galaxy rotation curves without invoking dark matter \cite{fr1,fr2,Mof96,mog,NAP,San84,Bro07}.
The light and weakly interacting $U$-boson is a
favorite candidate mediating the extra interaction
\cite{Fayet,Kri09,Zhu07}, and various new experiments in terrestrial
laboratories have been proposed to search for the $U$-boson, see,
e.g., Ref.~\cite{Yong13} and references therein. The search for evidence of the modified
gravity is at the forefront of research in several sub-fields of
natural sciences including geophysics, nuclear and particle physics,
as well as astrophysics and cosmology, see, e.g., Refs.\
\cite{Fujii71,Pea01,Ior02,Hoy03,Ark98,Long03,Kap07,Ior07a,Ior07b,Nes08,Kam08,Aza08,Ger10,Luc10,Bez10,Kli10}.
Various upper limits on the deviation from the ISL has been put
forward down to femtometer range using different experiments
\cite{Jun13}. These experiments have established a clear trend of
increased strength $\alpha$ at shorter length $\lambda$. In the
short range down to $\lambda \approx 10^{-15}-10^{-8}$ m, ground
state properties of nuclei, neutron-proton and neutron-lead
scattering data as well as the spectroscopy of antiproton atoms have
been used to set upper limits on the value of $g^{2}/\mu^{2}$ (or
equivalently the $|\alpha|~ vs ~\lambda$). Shown in the left window
of Fig. \ref{fig1} is a comparison of several recent constraints \cite{Jun13,Kam08,BARB75,POKO06,Nes08} in the $|\alpha|~
vs ~\lambda$ plane in the range of $\lambda \approx 10^{-15}-10^{-8}$ m obtained from nuclear laboratory experiments.
Extensive reviews on constraints at longer length scales are available in the literature, see, e.g.,  refs. \cite{Adel03,Fis99}.
We emphasize that below the upper limits shown here, the Yukawa term has no effect on properties of finite nuclei and nuclear scatterings. For example,
the upper limits on $|\alpha|$ in the femtometer scale were obtained by requiring explicitly that the binding energies and charge radii of nuclei do not deviate from
their experimental vales by more than $2\%$ \cite{Jun13}. As a reference, the straight lines are the $|\alpha|~ vs ~\lambda$ correlation for $g^{2}/\mu^{2}\approx 40-50$~GeV$^{-2}$.

Some justifications and cautions about using the Yukawa term to model modified gravity in neutron stars are in order here.
On one hand, the Yukawa term only appears in some of the modified gravity theories at the weak-field limit while neutron stars
have strong gravity fields. Indeed, in this sense it is highly questionable whether one can use the Fujii ansatz in studying neutron stars.
On the other hand, considered as the fifth force due to the exchange of the $U$-boson, a Yukawa term is possible independent of the strength of gravity.
As mentioned earlier, to our best knowledge, currently there is no fundamental reason to select one theory over others for describing the strong-field gravity as long as
the theory can explain all known facts from observations. Our reason of choosing the Fujii ansatz is mostly technical rather than physical.
To our best knowledge, the Yukawa-like non-Newtonian gravity is the only form that has its strength and
range parameters both limited using experimental data as illustrated in Fig. \ref{fig1} for the short-range part while the constraints for longer ranges can be found in the literature, see, e.g., refs. \cite{Kam08,BARB75,POKO06,Nes08}. Moreover, as we shall discuss in detail in the following, it is relatively easier to implement the Fujii ansatz and study its effects
on both static and oscillating properties of neutron stars, although its possible origins as a prediction of either the string theory or super-symmetric extension of the standard model may not be widely accepted.
There are many different kinds of modified gravity theories, and all of them have their own field equations governing static properties of neutron stars. Moreover, to determine the pulsating frequencies of neutron stars one has to use the first-order perturbations of the field equations. Such kinds of much more complicated studies are interesting and are being pursued by us and others using some of the modified gravity theories. However, they are beyond the scope of the present work.

Because of the degeneracy between the matter content and gravity in NSs,
a negative Yukawa term can be considered effectively either as an anti-gravity \cite{San84}
or a stiffening of the EOS \cite{Kri09}. According to Fujii \cite{Fuj2}, the Yukawa term in the boson exchange picture
is simply part of the matter system in general relativity. Therefore, effectively one can adjust the EOS
only while keep the same TOV equation for the static equilibrium for NSs.
For infinite nuclear matter, the extra energy density due to the Yukawa term is \cite{Long03,Kri09}
\begin{equation}\label{EDUB}
\varepsilon_ {_{\textrm{\scriptsize{UB}}}}= \frac{1}{2V}\int
\rho(\vec{x}_{1})\frac{g^{2}}{4\pi}\frac{e^{-\mu
r}}{r}\rho(\vec{x}_{2})d\vec{x}_{1}d\vec{x}_{2}=\frac{1}{2}\frac{g^{2}}{\mu^{2}}\rho^{2},
\end{equation}
where $V$ is the normalization volume, $\rho$ is the baryon number
density, and $r=|\vec{x}_{1}-\vec{x}_{2}|$. Assuming a constant
boson mass independent of the density, one obtains the additional
pressure
$P_{\textrm{\scriptsize{UB}}}=\frac{1}{2}\frac{g^{2}}{\mu^{2}}\rho^{2}$,
which is just equal to the additional energy density.
Including effects of the reduced gravity through the Yukawa term, the total pressure becomes
$P=P_{\textrm{\scriptsize{0}}}+P {_{\textrm{\scriptsize{UB}}}}$
where $P_ {0}$ is the nuclear pressure inside neutron stars.
Within this framework, the task of breaking the EOS-gravity degeneracy has effectively
become a question of separating effects of the $P_ {0}$  and $P_{\textrm{\scriptsize{UB}}}$.

As it was emphasized by Fujii \cite{Fuj2}, since the new vector boson
contributes to the nuclear matter EOS only through the combination
$g^{2}/\mu^{2}$, while both the coupling constant $g$ and the mass
$\mu$ of the light and weakly interacting bosons are small, the
value of $g^{2}/\mu^{2}$ can be large. On the other hand, by
comparing the $g^{2}/\mu^{2}$ value with that of the ordinary vector
boson $\omega$, Krivoruchenko et al. have pointed out that as long
as the $g^{2}/\mu^{2}$ value of the $U$-boson is less than about 200
GeV$^{-2}$ the internal structures of both finite nuclei and neutron
stars will not change \cite{Kri09}. However, global properties of
neutron stars can be significantly modified \cite{Kri09,Wen0911}.
One of the key characteristics of the Yukawa correction is its
composition dependence, unlike Einstein's gravity. Thus, ideally one
needs to use different coupling constants for various baryons
existing in neutron stars. Moreover, to our best knowledge, it is
unknown if there is any and what might be the form and strength of
the Yukawa term in the hadron-quark mixed phase and the pure quark
phase. Thus, instead of introducing more parameters, for the purpose
of this exploratory study, we assume that the
$P_{\textrm{\scriptsize{UB}}}$ term is an effective Yukawa
correction existing in all phases with the $g$ being an averaged
coupling constant.

For the EOSs of hybrid stars, we consider a quark core
covered by hyperons and leptons using models developed by us in several recent
studies of NS properties \cite{XCLM09a,XCLM09b,xu10,Lin1}. More
specifically, the quark matter is described by the MIT bag model
with reasonable parameters widely used in the literature
\cite{MIT1,MIT2}. The hyperonic EOS is modeled by using an extended
isospin- and momentum-dependent effective interaction (MDI)
\cite{Das03} for the baryon octet with parameters constrained by
empirical properties of symmetric nuclear matter, hyper-nuclei, and
heavy-ion reactions \cite{xu10}. In particular, the underlying EOS
of symmetric nuclear matter is constrained by comparing transport
model predictions with data on collective flow and kaon production
in high-energy heavy-ion collisions \cite{Pawel,LCK}. Moreover, the
nuclear symmetry energy $E_{sym}(\rho)$ with this interaction is
chosen to increase approximately linearly with density (i.e., the
MDI interaction with a symmetry energy parameter $x=0$ \cite{xu10})
in agreement with available constraints around and below the
saturation density \cite{Li13}. The Gibbs construction was adopted
to describe the hadron-quark phase transition \cite{Glen01}.
As known, the Gibbs construction requires that both the baryon and charge chemical potentials are in equilibrium, and it leads to a smooth phase transition. On the other hand, the Maxwell construction only requires the baryon chemical equilibrium, and the phase transition is first order. As mentioned in ref. \cite{xu10}, a more realistic EOS of the mixed phase can be obtained from the Wigner-Seitz cell calculation by taking into account the Coulomb and surface effects. The resulting EOS lies between those from the Gibbs and Maxwell constructions. The latter can thus be viewed as two extreme cases corresponding to certain values of the surface tension parameter in the Wigner-Seitz cell calculation. For the bag constant, the original MIT bag model gives a value of $B=55$ MeV/fm$^3$~\cite{MIT1}, while from lattice calculations it is about $B=210$ MeV/fm$^3$~\cite{Sat82}. Thus, the bag constant can be viewed as a free parameter in a reasonable range. In ref. \cite{xu10}, we compared the mass-radius relation with different values of the bag constant. It was shown that the phase transition happens at a higher density with a larger bag constant, while the maximum mass is insensitive to the value of the bag constant. To give a reasonable value of the transition density, we choose B$^{1/4}$=170 MeV (i.e., $B\approx 109$ MeV/fm$^3$) or 180 MeV (i.e., $B\approx 137$ MeV/fm$^3$) in the present study. The u and d quarks are taken as massless and the mass of the strange quark is set to be 150 MeV.

Similar to the previous work in the literature, the hybrid star is divided
into the liquid core, the inner crust, and the outer crust from the
center to surface. For the inner crust, a parameterized EOS of
$P_{\textrm{\scriptsize{0}}}=a+b \epsilon^{4/3}$ is adopted as in
Refs.\ \cite{XCLM09a,XCLM09b}. For the outer crust, the BPS
EOS~\cite{BPS} is adopted. As an example, shown in the right panel
of Fig.~\ref{fig1} are the EOSs for hybrid stars with MIT bag
constant $B^{1/4}=170$ MeV and $B^{1/4}=180$ MeV, respectively, with
and without the Yukawa contribution. The corresponding hadron-quark
mixed phase above/below the pure hadron/quark phase covers the
density range of $\rho/\rho_0=1.31$ to 6.56 and $\rho/\rho_0=2.19$
to 8.63, respectively. Including the Yukawa term the EOS stiffens as
the value of $g^{2}/\mu^{2}$ increases. It is seen that the two sets
of parameters, $B^{1/4}=170$ MeV and $g^{2}/\mu^{2}=$ 50~GeV$^{-2}$
or $B^{1/4}=180$ MeV and $g^{2}/\mu^{2}=$ 40~GeV$^{-2}$, lead to
approximately the same total pressure. As a reference, the MDI EOS
for the $npe\mu$ matter without hyperons and quarks is also shown.
In addition, for comparisons we also included in the right window of Fig.~\ref{fig1}
the soft and stiff EOSs for $npe$ matter constructed by Lattimer \cite{Lat12} and Hebeler et al. \cite{Heb13}.
These EOSs are constrained at sub-saturation densities by the EOS of pure neutron matter calculated by
a chiral effective field theory and at supra-saturation densities by the observed maximum masses of neutron stars.
It is interesting to note that our EOSs for the $npe\mu$ matter are very close to their EOSs for the $npe$ matter at low densities and are inside their soft-stiff EOS band at high densities.
We remak that our hadronic EOS, as shown in ref. \cite{Plamen}, is constrained by relativistic heavy-ion reaction data (flow and kaon production) in the density range of approximately $2-4.5\rho_0$ \cite{Pawel} which falls below the stiff EOS necessary to predict a $2.0M_{\odot}$ NS. However, the constraints from relativistic heavy-ion reactions were not considered in refs. \cite{Lat12,Heb13}. Interestingly, it is obvious that the soft-stiff EOS band from refs. \cite{Lat12,Heb13} leaves a large room for varying gravity theories at high densities.

\section{Breaking the EOS-gravity degeneracy with neutron star masses and pulsating frequencies}

\begin{figure}
\begin{center}
\includegraphics[width=8cm,height=6.65cm]{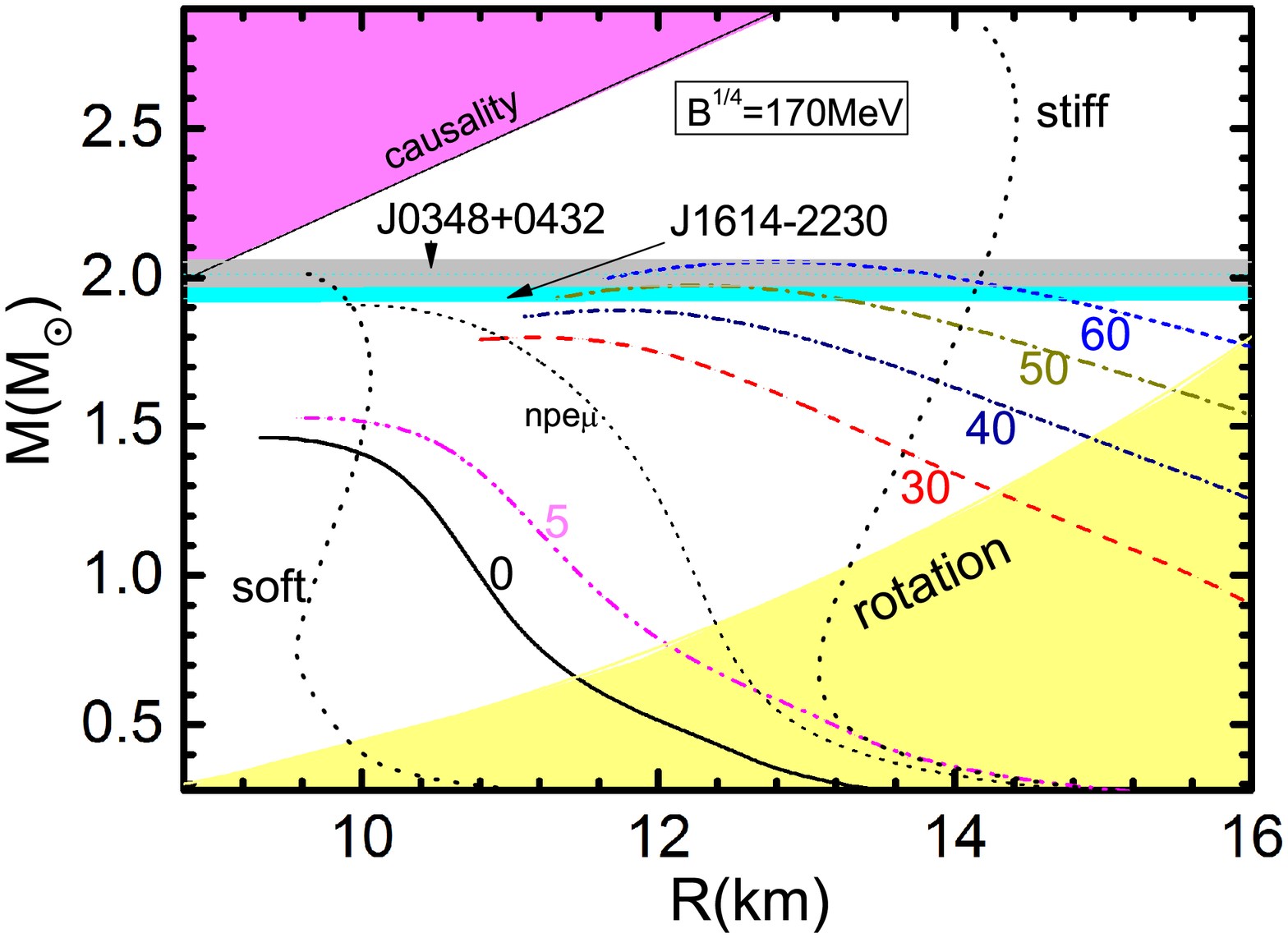}
\includegraphics[width=8cm,height=6.6cm]{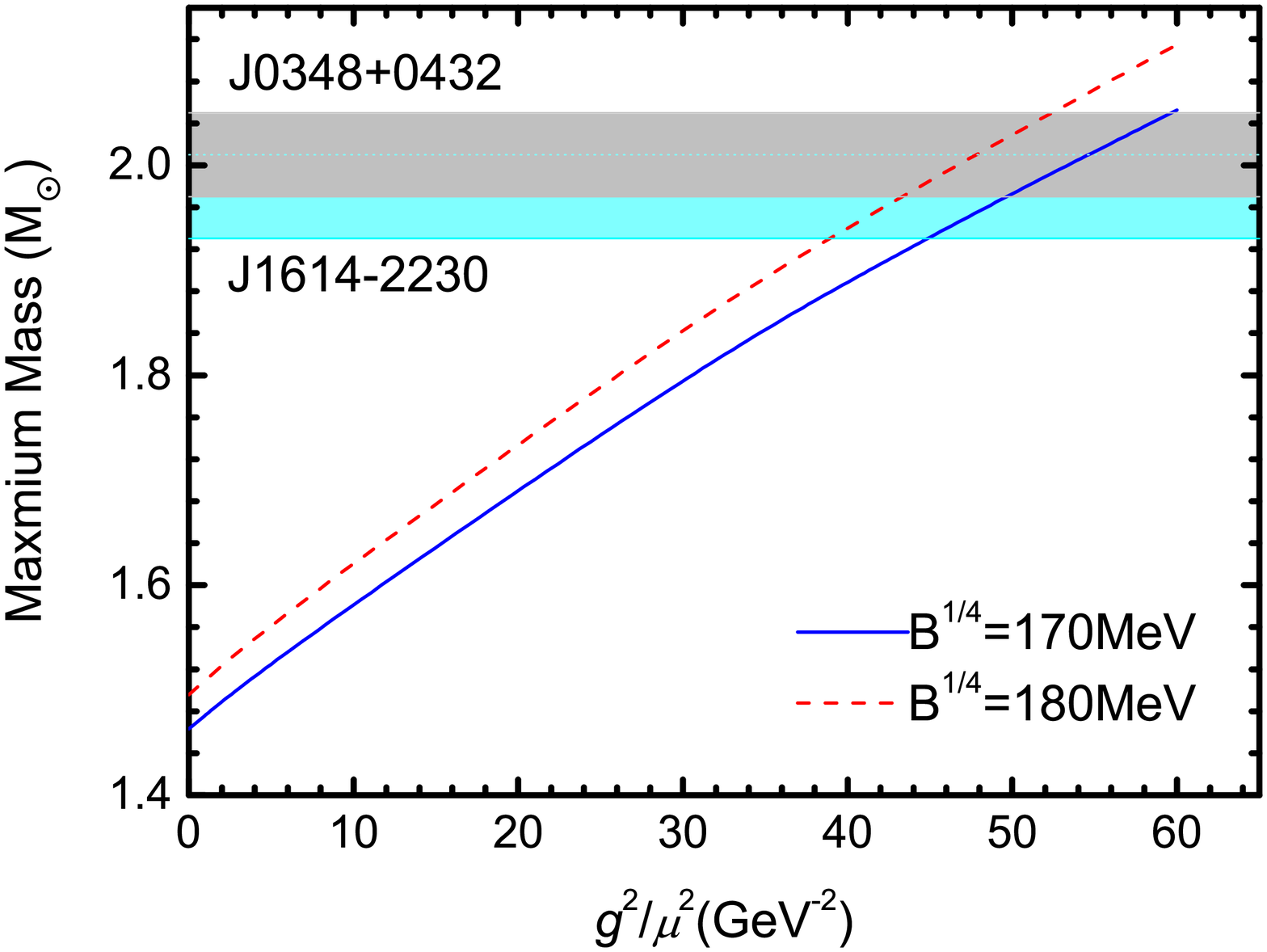}
\caption{\label{fig2}  (Color online) Left: The mass-radius relation
of static neutron stars with $B^{1/4}$=170 MeV and various values of
$g^{2}/\mu^{2}$ in units of GeV$^{-2}$ indicted using numbers above
the lines; results using the soft and stiff EOSs are taken from refs. \cite{Lat12,Heb13}.
Right: The maximum mass of neutron stars as a function of
$g^{2}/\mu^{2}$ with $B^{1/4}$=170 and 180 MeV, respectively.}
\end{center}
\end{figure}

Stellar oscillations of non-rotating compact stars in general relativity have
been extensively studied especially since the late sixties, see, e.g., Refs.
\cite{Thorne1,Thorne2,Thorne3,Thorne4,Thorne5,Lee1983,Lee1985,Chandrasekhar1990,Miniutti2003}.
Responding to the strong promise of detecting gravitational waves in the near future and the outstanding scientific opportunities provided by the advanced
detectors, the NS oscillating frequencies, especially the
fundamental ($f$) mode, the first few pressure ($p$) modes, and the
first pure space-time oscillation ($w_I$) have been studied with
much interest and fruitfully. In particular, using various model
EOSs the inverse problem of extracting information about the EOS of
supranuclear matter from future measurement of gravitational waves
from pulsating NSs in our own galaxy has been addressed in a number
of works, see, e.g., Refs. \cite{s1,Nil98,Kok,Ben05}. We examine here
simultaneously effects of the Yukawa-type non-Newtonian gravity and the
nuclear EOS on the frequencies of the $f$, $p_1$, $p_2$, and $w_I$
pulsating modes and the mass-radius correlation of NSs. In studying
the $f$, $p_1$, and $p_2$ modes, we used and compared numerical
algorithms developed in both Refs. \cite{Lee1983,Lee1985} and Refs.
\cite{Chandrasekhar1990,Miniutti2003}, and have reproduced the
results in \cite{Lee1983} and \cite{Benhar2004} within $1\%$. For
the $w_I$ mode, as in our previous work \cite{Lin1,Wen09}, we use
the continuous fraction method \cite{s4}.

Before trying to break the EOS-gravity degeneracy, we first examine
effects of the Yukawa-type non-Newtonian gravity on the mass-radius
correlation. Shown in the left panel of Fig. \ref{fig2} is the
mass-radius relation of hybrid stars with the bag constant
$B^{1/4}$=170 MeV and varying values of $g^{2}/\mu^{2}$. First of
all, without the Yukawa contribution (black solid line) the maximum
stellar mass supported is only about $1.46 M_{\odot}$. Including the
Yukawa term, as the EOSs are increasingly stiffened with larger
values of $g^{2}/\mu^{2}$, the maximum stellar mass increases. With
$g^{2}/\mu^{2}=50$ GeV$^{-2}$ the maximum mass of $1.97 M_{\odot}$
is just in the middle of the measured mass band of $1.97\pm0.04
M_{\odot}$ for PSR J1614-2230 \cite{Demo10}. The corresponding
radius is about 12.4 km. Moreover, one can see that the maximum
mass for the NS containing only the $npe\mu$ matter is similar to that
for hybrid stars calculated with $B^{1/4}$=170 MeV and $g^{2}/\mu^{2}=40$GeV$^{-2}$,
although the radius is smaller for the  $npe\mu$ star.  Compared to the latter, the softening of the EOS due to
the hyperonization and hadron-quark phase transition reduces,
while the effective stiffening of the EOS due to the reduced gravity increases
both the NS maximum mass and the corresponding radius.
To see more clearly relative effects of the bag
constant $B$ and the Yukawa term, shown in the right panel of
Fig.~\ref{fig2} are the maximum stellar masses as a function of
$g^{2}/\mu^{2}$ with $B^{1/4}=170$ MeV and $B^{1/4}=180$ MeV,
respectively. As expected, with $B^{1/4}=180$ MeV a smaller value of
$g^{2}/\mu^{2}=40~$GeV$^{-2}$ is needed to obtain a maximum mass
consistent with the observed masses of PSR J1614-2230 and
PSR J0348+0432. It is worth noting that the masses of these NSs
have ruled out many soft EOSs and a quark core. Many possible stiffening
mechanisms have been proposed during the last few years to
reproduce a maximum mass about $2 M_{\odot}$. The effective stiffening of the nuclear EOS due to
reduced gravity is one extreme but a possible mechanism. The results shown above indicate that measuring the NS masses alone
is insufficient to break the EOS-gravity degeneracy.
The simultaneous measurement of both the masses and radii can remedy the situation.
Unfortunately, it is currently under hot debate whether the radii are
around 8-10 km \cite{Ozel,Bob}, 10-12 km \cite{Ste10}, or larger than 14 km \cite{Sul11}
because of the well-known difficulties and model dependence in extracting the radii of neutron stars.
Thus, alternative observables should be investigated.

\begin{figure}
\includegraphics[width=7cm]{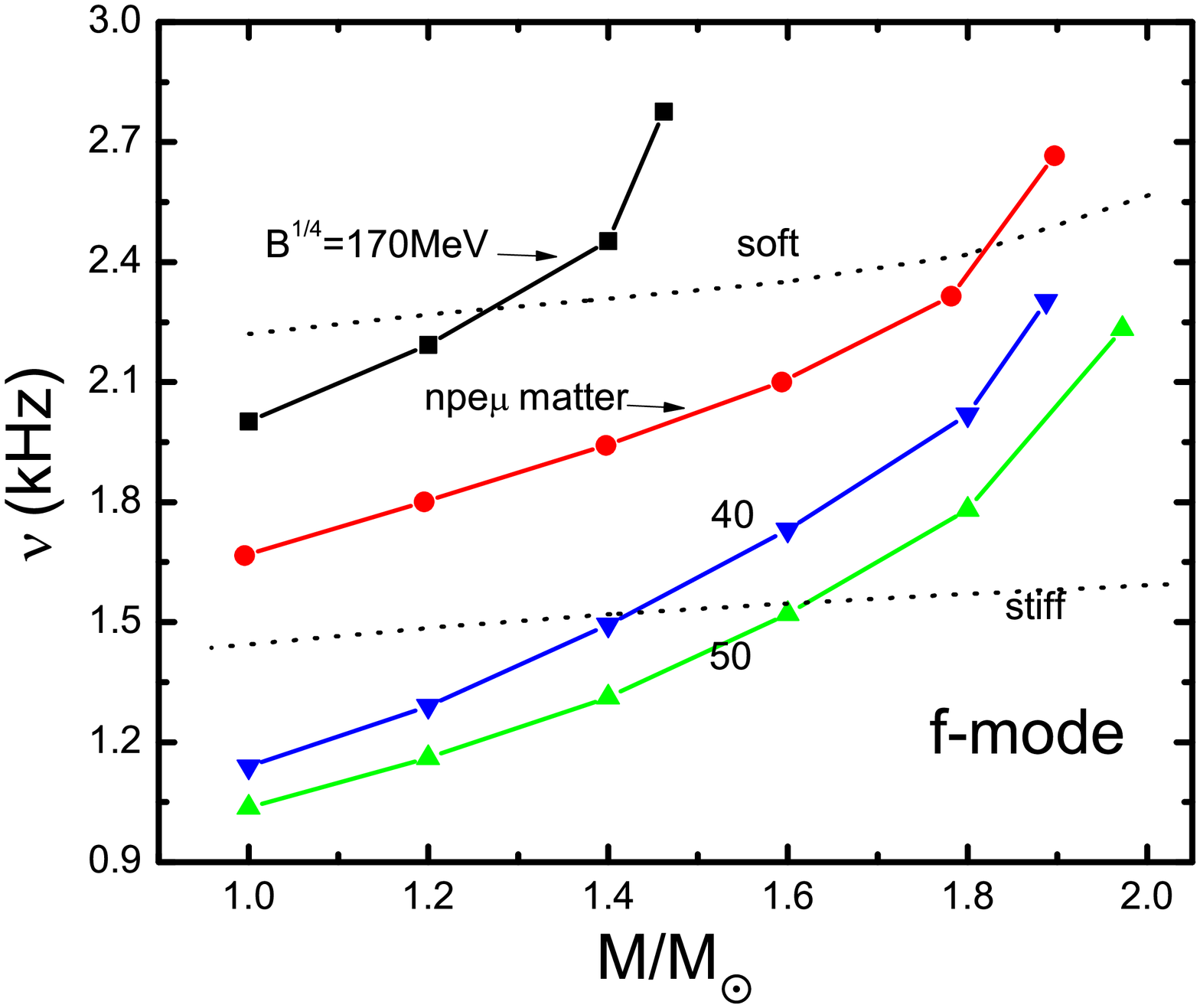}
\includegraphics[width=7cm]{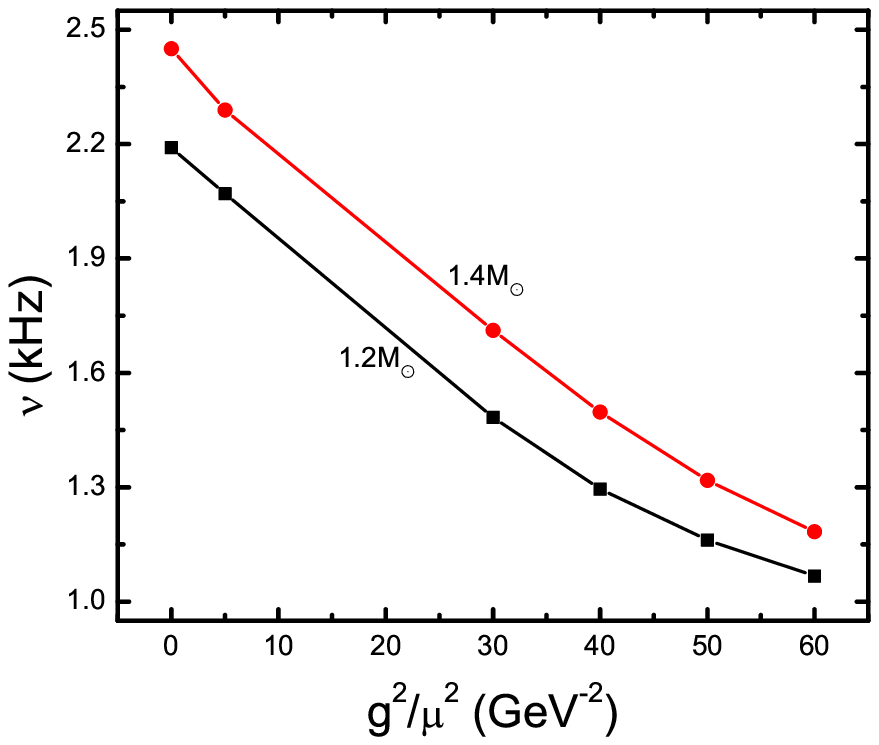}
\caption{\label{fig3}  (Color online) Left: The $f$-mode frequency
versus the NS mass. The upper two curves are results without
including the Yukawa potential while the lower two are obtained with
$g^2/\mu^2=40$ and $50$ GeV$^{-2}$ (for hybrid stars with $B^{1/4}$=170 MeV). The soft-stiff results are obtained using the EOSs of refs. \cite{Lat12,Heb13}; Right: The $f$-mode frequency
versus the strength of Yukawa-type non-Newtonian gravity $g^2/\mu^2$
for neutron stars of mass $1.4 M_{\odot}$ and $1.2 M_{\odot}$.}
\end{figure}

\begin{figure}
\includegraphics[width=0.8\textwidth]{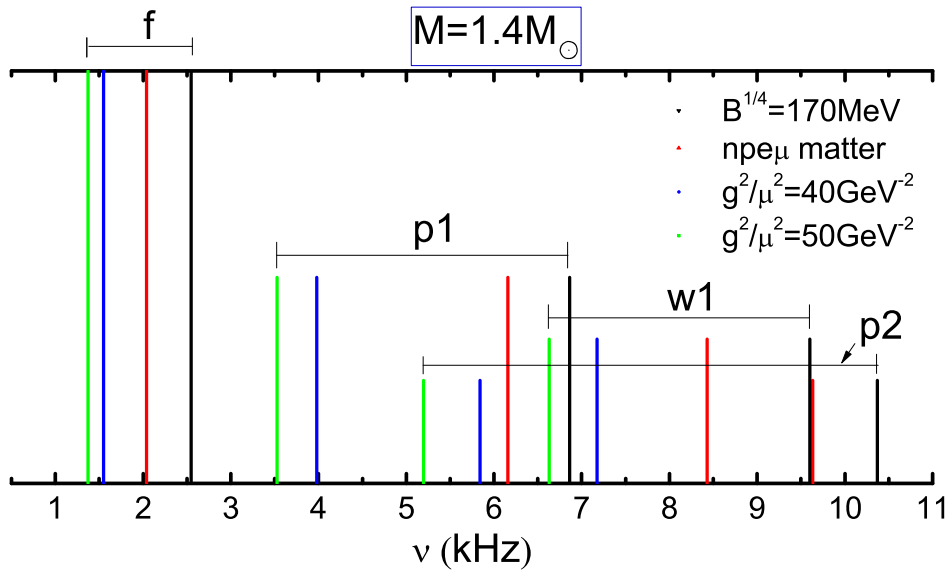}
\caption{\label{fig4} (Color online) Comparison of the $f$, $p_1$,
$p_2$, and $w_I$ pulsating frequencies for a neutron star of mass
$1.4 M_{\odot}$. The vertical scale is arbitrary.}
\end{figure}

We now turn to the pulsating frequencies of NSs and their dependence
on both the nuclear EOS and the strength $g^2/\mu^2$ of the Yukawa-type non-Newtonian
gravity. Shown in the left panel of Fig. \ref{fig3} is the $f$-mode
frequency versus NS mass. The upper two curves are results without
including the Yukawa potential while the lower two are obtained with
$g^2/\mu^2=40$ and $50$ GeV$^{-2}$ (for hybrid stars with $B^{1/4}$=170 MeV) necessary to obtain a maximum mass
of about $2 M_{\odot}$. Generally, the $f$-mode frequency is higher
with a softer EOS consistent with findings in earlier studies, see, e.g., Refs. \cite{Kok,Ben05}.
Results using the soft and stiff EOSs of refs. \cite{Lat12,Heb13} are also shown. It is interesting to see that
the soft-stiff band only allows possible variation of gravity above about $1.4 M_{\odot}$. Within the band, the frequency of the f-mode is sensitive to the variation of
$g^2/\mu^2$. With the Yukawa-type non-Newtonian gravity the
$f$-mode frequency decreases significantly while the maximum mass
increases. It is also interesting to see that although the maximum
masses are almost the same for the NS containing $npe\mu$  matter only and the hybrid star with $B^{1/4}$=170 MeV and
$g^{2}/\mu^{2}=40$ GeV$^{-2}$, their $f$-mode frequencies are rather different.
To be more quantitative about effects of the non-Newtonian gravity, shown in the right
panel of Fig. \ref{fig3} are the $f$-mode frequency versus the strength
$g^2/\mu^2$ for NSs of mass $1.4 M_{\odot}$ and $1.2 M_{\odot}$, respectively. Combining the results
shown in both panels of Fig. \ref{fig3}, it is obvious that if both
the mass and the $f$-mode frequency are measured, the EOS-gravity
degeneracy is readily broken. Similar gravity effects were found on
the $p_1$, $p_2$, and $w_I$-modes. Depending on the frequency, it
requires different amount of energy to excite various pulsating
modes of NSs. Shown in Fig. \ref{fig4} is a comparison of the $f$,
$p_1$, $p_2$, and $w_I$ pulsating frequencies for a neutron star of
mass $1.4 M_{\odot}$ with or without the Yukawa term. While effects
of the Yukawa term on the $p_1$ and $p_2$ modes are relatively
larger, these modes are harder to excite compared to the $f$ mode.
We notice also that frequencies of the $f$ and $p_1$ modes are
within reach by the advanced gravitational detectors available
although not at their peak sensitivities, while the detection of the $w_I$
mode requires new detectors capable of measuring gravitational at high frequencies.
Thus, the $f$-mode is currently the most promising one for breaking the EOS-gravity degeneracy. The pulsating
modes also have their respective damping times characterized by the
imaginary parts of their complex frequencies. We have also studied
effects of both the nuclear EOS and gravity on the damping times, but did
not find anything particularly useful for the purpose of this work.

\section{Conclusions}

Both the EOS of super-dense nuclear matter and the strong-field gravity
are poorly known and their determinations are at the forefronts of nuclear
physics and astrophysics, respectively. A comprehensive and simultaneous
understanding of both of them are required to interpret unambiguously observed
properties of neutron stars because of the EOS-gravity degeneracy.
To break the degeneracy, at least two independent observables are required.
In this work, to explore potentially useful observables for breaking the EOS-gravity degeneracy,
we investigated effects of the nuclear EOS and Yukawa-type non-Newtonian Gravity on the mass-radius correlation,
the $f$-mode, $p$-mode, and $w$-mode stellar oscillations of neutron stars.
We found that all the NS pulsating frequencies are significantly lowered while their masses
are increased by the Yukawa term for a given nuclear EOS. A simultaneous
measurement of the NS mass and the frequency of one of its pulsating
modes, preferably the $f$ mode, can thus readily break the EOS-gravity degeneracy.
With the already available high-quality x-ray observational data of neutron stars and the various ongoing upgrades of
advanced gravitational wave detectors, we hope such a combination of observables will be possible in the near future.

\section{Acknowledgement}

We would like to thank F. Fattoyev, W. Z. Jiang, and W. G. Newton
for useful discussions. This work was supported in part by the US
National Science Foundation under Grant No. PHY-1068022, the US
National Aeronautics and Space Administration under grant NNX11AC41G
issued through the Science Mission Directorate, the CUSTIPEN
(China-U.S. Theory Institute for Physics with Exotic Nuclei) under
US DOE grant number DE-FG02-13ER42025, the National Natural Science
Foundation of China under Grant No. 10947023, 11275125, 11135011,
11175085, 11235001, 11035001, 11275073 and 11320101004, the Shanghai Rising-Star
Program under Grant No. 11QH1401100, the ``Shu Guang" project
supported by Shanghai Municipal Education Commission and Shanghai
Education Development Foundation, the Program for Professor of
Special Appointment (Eastern Scholar) at Shanghai Institutions of
Higher Learning, the Science and Technology Commission of Shanghai
Municipality (11DZ2260700), the ``100-talent plan" of Shanghai
Institute of Applied Physics under Grant No. Y290061011 from the
Chinese Academy of Sciences,  Shanghai Pujiang Program under Grant No. 13PJ1410600,
and the Chinese Fundamental Research Funds for the Central Universities (Grant No. 2014ZG0036).

\end{document}